\documentstyle[12pt]{article}
\newcommand{\lea}{\raisebox{-.3ex}{\small $ \
\stackrel{\textstyle <}{\sim} $ }}

\begin{document}

\begin{center}
{\LARGE\bf Nucleon-Nucleon Potentials in Comparison: Physics or
Polemics?\footnote{Dedicated to Tom Kuo on the occasion of his
60th birthday.}}
\\
\vspace*{.5cm}
{\large\sc R. Machleidt\footnote{Invited talk presented at `Realistic
Nuclear Structure', a conference to mark the 60th Birthday of T. T. S.
Kuo, May 1992, Stony Brook, New York.} and G. Q. Li}
\\
{\it Department of Physics, University of Idaho,
 Moscow, ID 83843, U.S.A.}
\end{center}
\vspace*{.5cm}

\begin{abstract}
Guided by history, we review the major developments concerning
realistic nucleon-nucleon (NN) potentials
since the pioneering work by Kuo and Brown on the
effective nuclear interaction.
Our main emphasis is on the physics underlying various models
for the NN interaction developed over the past quarter-century.
We comment briefly
on how to test the quantitative nature of nuclear potentials
properly.
A correct calculation (performed by independent researchers)
of the $\chi^2$/datum for the fit of the
world NN data yields
5.1, 3.7, and 1.9 for the Nijmegen, Paris, and Bonn potential,
respectively.
Finally, we also discuss in detail the relevance of the on- and off-shell
properties of NN potentials for microscopic nuclear structure calculations.
\end{abstract}

\section{Introduction}
To most of us, Tom Kuo is best known for his
seminal work on the effective interaction of two valence nucleons
in nuclei (`Kuo-Brown matrix elements') published in 1966~\cite{KB66}.
One of the most fundamental goals of theoretical nuclear physics
is to understand atomic nuclei in terms of the basic
nucleon-nucleon (NN) interaction. Tom and Gerry's work of 1966 was
the first successful step towards this goal.
Microscopic nuclear structure has essentially
two ingredients:
many-body theory and the nuclear potential.
During the past 25 years, there has been progress in both of these fields.
Tom Kuo has worked consistently on the improvement of the many-body
approaches appropriate for nuclear structure
problems~\cite{KLR71,SKS83}.
On the other hand, there have also been substantial advances in
our understanding of the NN interaction since 1966, when the
Hamada-Johnston potential~\cite{HJ62} was the only available
quantitative NN potential.

Therefore,
we are happy to take this opportunity to give an overview
of the progress made in the field of realistic two-nucleon
potentials
in the past quarter-century.
It is not our intention to mention and analyse all NN potentials that
have appeared on the market ever since
 the work of Hamada/Johnston~\cite{HJ62}
and Reid~\cite{Rei68}.\footnote{A more detailed account of this can be
found in section~2 of ref.~\cite{Mac89}.}
Here, our focus will be on the physics underlying realistic models
for the nuclear potential. Most transparently, this physics has evolved
in three stages, and we will give essentially only one representative
example for each stage (section~2).

We do so to avoid confusing the reader with too many technical details
that may distract from the interesting physics involved.
However, for reasons of fairness, we like to stress that
our sub-selection is essentially accidental, and that there are many
other, equally representative examples for each stage of the
development of theoretical nuclear potentials. In particular, we
mention the potential by Tourreil, Rouben, and Sprung~\cite{TRS75},
the Argonne $V_{14}$ potential~\cite{WSA84}, and the more recent models
developed at Bochum~\cite{Bochum}, Williamsburg~\cite{GOH92}, and
Hamburg~\cite{Hamburg}.

With an eye on the topic of this conference, we like to restrict
our review to quantitative NN potentials suitable for
application in nuclear structure. Therefore, we will not discuss
the numerous and interesting recent attempts of deriving the NN interaction
from models for low-energy QCD.

For microscopic nuclear structure work, it is important that the NN
potential applied is able to reproduce
 the known facts about the two-nucleon system
correctly.
Therefore, in section~3, we will make some comments on how to test
the quantitative nature of a two-nucleon potential properly.
Most relevant for this conference is the question to which extent
nuclear structure results depend on the kind of potential
used as input in the calculations. We will discuss this in
section~4. Our contribution finishes with summary and conclusions given in
section~5.

\section{Physics}
The physics underlying different meson-theory based
models for the NN interaction
will be explained in this section. This is done most clearly
by following the historical path.

\subsection{The one-boson-exchange model and the Nijmegen potential}

The first
quantitative meson-theoretic models for the NN interaction
were the one-boson-exchange potentials (OBEP).
They emerged after the experimental
discovery of heavy mesons in the early 1960's.
In general, about
six non-strange bosons with masses below 1 GeV are taken into account
(cf.\ fig.~1): the pseudoscalar mesons $\pi (138)$ and $\eta (549)$,
the vector mesons $\rho (769)$ and $\omega (783)$, and two scalar bosons
$\delta (983)$ and $\sigma (\approx 550)$. The first particle
in each group
is isovector while the second is isoscalar.
The pion provides the tensor force, which is reduced at short range
by the $\rho$ meson.
The $\omega$ creates the spin-orbit force and the short-range repulsion,
and the $\sigma$ is responsible for the intermediate-range attraction.
Thus, it is easy to understand why a model which includes the above four
mesons can reproduce the major properties of
the nuclear force.\footnote{The
interested reader can find a detailed and pedagogical
introduction into the OBE model in
section~3 and 4 of ref.~\cite{Mac89}.}

A classic example for an OBEP is the
Bryan-Scott (B-S) potential published in 1969~\cite{BS69}.
Since it is suggestive to think of a potential as a function of $r$
(where $r$ denotes the distance between the centers of the two interacting
nucleons), the OBEP's of the 1960's where represented as
local $r$-space potentials.
To reduce the original one-meson-exchange Feyman amplitudes
to
such a simple form, drastic approximations have to be applied.
The usual method is to expand the amplitude in terms of $p^2/M$
and keep only terms up to first order
(and, in many cases, not even all of them).
Commonly, this is called the
{\it non-relativistic OBEP}.
Besides the suggestive character of a local function of $r$,
such potentials are easy to apply in $r$-space calculations.
However, the original potential (Feynman amplitudes) is non-local,
and, thus, has a very different off-shell behaviour than its
local approximation. Though this does not play a great role in two-nucleon
scattering, it becomes important when the potential
is applied to the nuclear few- and many-body problem.
In fact, it turns out that the original non-local potential
leads to much better predictions in nuclear struture than the
local approximation.
We will discuss
this in detail in section~4.

Historically, one must understand that
after the failure of the pion theories in the 1950's,
the one-boson-exchange (OBE) model was considered a great
success in the 1960's.
However, there are conceptual and quantitative problems with this model.

A principal deficiency of the OBE model is the fact that it has to
introduce a
scalar-isoscalar $\sigma $-boson in the mass range
500--700 MeV, the existence of which is
not supported by any experimental evidence. Furthermore, the model is
restricted to single exchanges of bosons that are `laddered' in an unitarizing
equation. Thus, irreducible multi-meson exchanges, which may be quite
sizable (see below), are neglected.

Quantitatively, a major drawback of the {\it non-relativistic}
 OBE model is its failure to
describe certain partial waves correctly.
In fig.~2a and b, we show phase shifts for the $^1P_1$ and $^3D_2$ state,
respectively. It is clearly seen that the Bryan-Scott non-relativistic OBE
potential
(B--S, long dashes)
predicts these phases substantially above the data.

An important progress of the 1970's  has been the development of the
{\it relativistic OBEP}~\cite{ROBEP}. In this model, the full, relativistic
Feynman amplitudes for the various one-boson-exchanges
are used to define the potential. These non-local expressions do not
pose any numerical problems  when used in momentum space.\footnote{
In fact, in momentum space, the application of a non-local
potential is numerically
as easy as using the momentum-space representation of a local
potential.}
The quantitative deficiencies of the non-relativistic OBEP disappear
immediately when the non-simplified, relativistic, non-local
OBE amplitudes are used.

The {\it Nijmegen potential}~\cite{NRS78}, published in 1978,
 is a non-relativistic $r$-space OBEP.
As a late representative of this model, it is one of the most
sophisticated examples of its kind.
It includes all non-strange mesons of the pseudoscalar, vector, and
scalar nonet.
Thus, besides the six mesons mentioned above,
the $\eta'(958)$, $\phi(1020)$, and
$S^*(993)$ are taken into account.
The model also includes the dominant J=0 parts of the Pomeron and
tensor ($f$, $f'$, $A_2$)
trajectories, which essentially lead to repulsive
central Gaussian potentials.
For the pomeron (which from
today's point of view may be considered as a multi-gluon exchange)
a mass of 308 MeV is assumed.
At the meson-nucleon vertices, a cutoff (form factor)
of exponential shape is applied
(most OBEP use cutoffs of monopole type).
However, this potential is still defined in terms
of the non-relativistic local approximations to the OBE amplitudes.
Therefore, it shows exactly the same problems
as its ten years older counterpart, the Bryan-Scott potential, cf.\ fig.~2:
the Nijmegen potential fails to predict the $^1P_1$ and $^3D_2$ phase shifts
correctly almost to the same extent as
the Bryan-Scott potential.
In addition, the Nijmegen potential overpredicts the $^3D_3$ phase shifts
by more than a factor of two (fig.~2c), a problem the models of the 1960's
did not have.  The inclusion of more bosons did obviously
not improve the fit of the data.
In fact, in the case of the $^3D_3$,
there is a dramatic deterioration as compared to earlier models
(cf.\ fig.~2c).

Spin observables
of neutron-proton scattering are quite sensitive to the
$^3D_2$ and $^3D_3$ partial waves.
Therefore, a quantitative description of these waves
is important. In fig.~3
we show the spin correlation parameter $C_{NN}$ at 181 MeV as measured
at the Indiana Cyclotron~\cite{Indiana} and at 220 MeV
with the data from TRIUMF~\cite{TRIUMF}.
Failures to reproduce this observable can be clearly
traced to the $^3D_2$ and $^3D_3$ (cf.~fig.~2b and c).

\subsection{The $2 \pi$-exchange and the Paris potential}

In the 1970's, work on the meson theory of the nuclear force focused on the
$2\pi$-exchange contribution to the NN interaction to replace the fictitious
$\sigma$-boson. One way to calculate these contributions is by means
of dispersion relations (fig.~4). In this approach, one assumes that the total
$2\pi$ exchange contribution to the NN interaction, fig.~4a, can be analysed
in terms of two `halves' (fig.~4b). The hatched oval stands for all possible
processes involving a pion and a nucleon. More explicitly, this is
shown in figs.~4c-e. The amplitude fig.~4c is calculated from empirical input
of $\pi N$ and $\pi\pi$ scattering. Around 1970, many groups throughout the
world were involved in this approach; we mention here, in particular,
the Stony Brook~\cite{JRV75}
and the Paris group~\cite{Vin73}.
These groups could show that the intermediate-range part of the nuclear force
is, indeed, decribed correctly by the $2 \pi$-exchange
as obtained from dispersion
integrals (see fig.~7, below).

To construct a complete potential, the Stony Brook as well as the Paris
group complemented their $2 \pi$-exchange contribution by one-pion-exchange
(OPE)
and $\omega$ exchange. In addition to this, the Paris potential contains a
phenomenological short-range potential for $r\lea 1.5$ fm.
For some components of the Paris potential, this short-range phenomenology
changes the original theoretic potential considerably, see fig.~5.

In the final version of the Paris potential,
 also known as the parametrized
Paris potential~\cite{Lac80},
each component (there is a total of 14 components,
7 for each isospin) is parametrized in terms of 12 local Yukawa
functions of multiples of the pion mass. This introduces a very large
number of parameters, namely $14 \times 12 = 168$. Not all 168 parameters
are free. The various components of the potential are required to vanish
at $r=0$ (implying 22 constraints~\cite{Lac80}).
One parameter in each component
is the $\pi NN$ coupling constant, which may be taken from other
sources (e.~g., $\pi N$ scattering). The $2 \pi$-exchange
contribution is derived from dispersion theory.
The range of this contribution is typically equivalent to an
exchanged  mass of 300--800 MeV or,
in terms of multiples of the pion mass ($m_\pi$),
$2 m_\pi - 6 m_\pi$. Thus,
in parametrized from, this contribution requires
5 parameters per component (implying a total of
$5\times 14 = 70$ parameters) that are fixed by theory and not
varied during the best-fit of NN data. In summary, one may assume
that about 100 parameters are constrained by the theory
underlying the Paris potential; this leaves about 60 free
parameters that are used to optimize the fit of the NN data.
These parameters  mainly determine the short-range part of the
Paris potential
($r\lea 1.5$ fm), cf.~fig.~5.

Another way to estimate the number of free parameters
is to start from the observation that deviations
from the original theoretical potential
by the
best-fit potential
occur for $r \lea 1.5$ fm
(cf.\ fig.~5). This range is equivalent to about
6$m_\pi$ and more. Thus, the Yukawas with six and more pion masses
are affected by the best-fit procedure. Since there are 12 Yukawas,
there are seven larger or equal 6;
this applies to the 14 components of the potential
separately, yielding $7\times 14 = 98$ parameters
from which the above mentioned 22 constraints have to be subtracted;
the result is 76 fit parameters.
Thus, our above estimate of 60 free parameters is on the conservative side.

There are advantages and disadvantages to a large number of fit parameters.
The obvious advantage is that with many parameters a good fit
of the NN scattering data can easily be obtained;
the Paris potential fits the NN data well (see table~1, below).
However, from a more basic point of view, a large
number of phenomenological parameters is discomforting:
the theory, underlying the potential, is obscured
and, thus, cannot be put to a real test.

To explain the latter point in more datail:
For the quantitative description  of the low-energy
NN data ($E_{lab}\lea 300$ MeV), the S-, P-, and D-waves are crucial.
A good $\chi^2$ for the fit of the NN data by a potential is due
mainly to a good reproduction of the S-, -P, and D-wave phase shifts.
These waves are essentially determined by the
short-range potential ($r\lea 1.5$ fm).
Thus, in the Paris potential these
lower partial waves are chiefly a product
of the  phenomenological part of the potential.
Consequently, the good fit of the data
by the Paris potential has primarily to do with its phenomenological
character at short range and not with the underlying meson theory.
The theory, the Paris potential is based upon, describes the peripheral
partial waves ($L\geq $4) well (cf.\ fig.~7, below). In fact,
the consideration of partial waves with $L\geq 4$
is the real test of the theory underlying the Paris potential,
and it passes this test well.
However, for the low-energy NN observables, the peripheral
waves are of little significance, the bulk is provides
by
$L\leq 3$.

Thus, by comparing the experimental NN scattering data with the
predictions by the Paris potential, meson-theory is not put to a test.
This comparison tests the phenomenological potential
which is, indeed, consistent with the data.
The satisfactory $\chi^2$ of the fit of the data by the Paris potential
(cf.\ table~1) cannot be used as a proof
that meson-theory is correct for the
low-energy NN system.
This is unfortunate, since it is this type of information
that one would like to draw from a low-energy,
meson-theory based potential.
In sect.~2.4, below, we will explain how to put meson-theory
to a true test.

In microscopic nuclear structure calculations, the off-shell
behavior of the NN potential is important (see sect.~4
for a detailed discussion).
The fit of NN potentials to two-nucleon data
fixes them on-shell.
The off-shell behavior cannot, by principal,
be extracted from two-body data.
Theory could
determine the off-shell nature of the potential.
However, not any theory can do that. Dispersion theory relates
observables (equivalent to on-shell $T$-matrices) to observables;
e.~g., $\pi N$ to $NN$.
Thus, dispersion theory cannot, by principle,
provide any off-shell information.
The Paris potential is based upon dispersion theory; thus, the off-shell
behavior of this potential is not determined by the underlying
theory.
On the other hand, every potential does have an off-shell behavior.
When undetermined by theory, then the off-shell behavior is a
silent by-product of the parametrization chosen to fit the
on-shell $T$-matrix, with which the potential is identified,
by definition.
In summary, due to its basis in dispersion theory, the
off-shell behaviour of the Paris potential is
not derived on theoretical grounds.
This is a serious drawback when it comes to the question of how to
interpret nuclear structure results obtained by
applying the Paris potential.

\subsection{The field-theoretic approach to the $2 \pi$-exchange}

In a field-theoretic picture,
the interaction between mesons and baryons can be described by
effective Lagrangians.
The NN interaction can then be derived in terms of field-theoretic
perturbation theory.
The lowest order (that is, the second order in
terms of meson-baryon interactions) are
the one-boson-exchange diagrams (fig.~1), which are easy to calculate.

More difficult (and more numerous) are the irreducible two-meson-exchange
(or fourth order) diagrams.
It is reasonable to start
with the contributions of longest range. These are the graphs that
exchange two pions. A field-theoretic model for this $2\pi$-exchange
contribution to the NN interaction is shown in fig.~6.
Naively, one would expect only the two diagrams in the first row
of fig.~6.
However, there are two complications that need to be taken into account:
meson-nucleon resonances and meson-meson scattering.
The lowest $\pi N$ resonance, the so-called $\Delta$
isobar with a mass of 1232 MeV gives rise
to the diagrams in the second and third row of fig.~6. The last two rows
include $\pi \pi$ interactions. While two pions in relative $P$-wave
form a resonance (the $\rho$ meson),
 there is no proper resonance in $\pi\pi-S$-wave below 900 MeV.
However, there are strong correlations in $S$-wave at low energies.
Durso {\it et al.}~\cite{DJV80} have shown that these correlations
can be described in terms of a broad mass distribution of about
$600\pm 260$ MeV, which in turn can be approximated by
a zero-width scalar-isoscalar boson of mass 550 MeV~\cite{MHE87}.

At this point, two approaches are available for calculating
the $2\pi$-exchange
contribution to the NN interaction:
dispersion theory (Paris~\cite{Vin73}) and field theory (Bonn~\cite{MHE87}).
Let us now compare the predictions by the two approaches
with each other as well as with the data.
For this purpose, it is appropriate to look into
the peripheral partial waves of NN scattering.
In figure~7, the curve ``BONN'' represents the predictions by the
field-theoretic model of fig.~6; the dotted curve labeled ``P'73''
is the original Paris result~\cite{Vin73} as obtained from
dispersion theory, while the dotted curve ``P'80'' is the result
from the parametrized Paris potential~\cite{Lac80} (all curves also
contain one-pion and $\omega$ exchange).
The agreement between the two theoretical curves and the data is
satisfactory.
However, there are some substantial deviations
(particularly in $^3F_4$) between the theoretical results by the Paris
group and their parametrized version.
This is puzzling, because we are dealing here with the long-
and intermediate-range
potential (that is not altered in the parametrization process).

\subsection{$\pi \rho$-contributions and the Bonn potential}
As demonstrated in the previous subsection, a model consisting of
$\pi+2\pi+\omega$
describes the peripheral partial waves quite satisfactory.
However, when proceeding to lower partial waves
(equivalent to shorter internucleonic distances),
this model generates too much attraction.
This is true for the dispersion-theoretic result (Paris)
as well as the field-theoretic one (Bonn).
For D-waves, this is seen in fig.~8 by the curves labelled
`$2\pi$ Paris' and `$2\pi$ Bonn'.
Obviously, further measures have to be taken to arrive at a quantitative
model for the NN interaction.
It is at this point that the philosophies of the Bonn and Paris group diverge.

The Paris group decided to give up meson theory at this stage and to
describe everything that is still missing by phenomenology.
Thus, the S-,  P-, and largely also the D-waves
in the Paris potential are essentially fitted by phenomenology.
The difference between the `$2\pi$ Paris' and the `Paris' curve
in fig.~8 is the
effect of the short-range phenomenological
Paris potential choosen such as to fit the empirical phase shifts.

In contrast, the Bonn group continued to consider further irreducible
two-meson exchanges. The next set of
diagrams to be considered are the exchanges
of $\pi$ and $\rho$ (fig.~9).
Notice that these diagrams are analogous to the upper six
$2\pi$-exchange diagrams of fig.~6.
The effect of the $\pi\rho$ contributions in D-waves is seen
in fig.~8 by comparing
the curve `$2\pi$ Bonn' with `$2\pi + \pi\rho$ Bonn'.
Clearly, these contributions very accurately take care  of the
discrepancies that were remaining between theory and experiment.

Even more remarkable is the effect of the $\pi\rho$ contributions
in S- and P-waves. We show this in fig.~10 where the dashed
line represents the phase shift prediction of the $\pi+2\pi+\omega$
model while the solid line includes the $\pi\rho$ diagrams.

In figs.~8 and 10, we have demonstrated clearly that the $\pi\rho$
contributions provide the short-range repulsion
which was still missing. It is important to note that the $\pi\rho$
contributions have only one free parameter, namely the cutoff for
the $\rho N \Delta$ vertex. The other parameters involved occur also
in other parts of the model and were fixed before (like the $\pi NN$
and $\rho NN$ coupling constants and cutoff parameters).
Notice also that the $\pi N\Delta$
and $\rho N \Delta$ coupling constants are not free parameters, since
they are related to the corresponding $NN$ coupling constants by
$SU(3)$.

In the light of the $\pi\rho$ results shown, the bad description
of some P- and D-waves by
some NN potentials (cf.~fig.~2, above)
may be simply understood as a lack of the $\pi\rho$
contributions.

In summary, a proper meson-theory for the NN interaction should
include the diagrams of irreducible $\pi\rho$ exchange.
This contribution has only one free parameter and
makes comprehensive short-range phenomenology
unnecessary. Thus, meson theory can be truly tested
in the low-energy NN system.

In the 1970's and 80's,
a model for the NN interaction was developed at the University of
Bonn. This model
consists of single $\pi$, $\omega$, and $\delta$ exchange, the $2\pi$
model shown in fig.~6, and the $\pi\rho$ diagrams of fig.~9,
as well as
a few more irreducible $3\pi$ and $4\pi$  diagrams (which are not very
important, but indicate convergence of the diagrammatic expansion).
This quasi-potential
has become known as the `Bonn full model'~\cite{MHE87}.
It has 12 parameters which are the coupling constants and cutoff masses
of the meson-nucleon vertices involved.
With a reasonable choice for these parameters, a very satisfactory
description of the NN observables up  about 300 MeV is achieved
(see table~1).
Since the goal of the Bonn model was to put meson theory
at a real test,
no attempt
was ever made to minimize the $\chi ^2$ of the fit of the NN data.
Nevertheless, the Bonn full model shows the smallest $\chi^2$
for the fit of the NN data (cf.~table~1).

\subsection{Summary}
In table 1, we give a summary and an overview of the theoretical
input of some meson-theoretic NN models discussed in the previous sections.
Moreover, this table also lists the $\chi^2$/datum (as calculated by
independent researchers~\cite{SAID}) for the fit of
the relevant
world NN data, which is 5.12, 3.71, and 1.90 for the Nijmegen~\cite{NRS78},
Paris~\cite{Lac80}, and Bonn~\cite{MHE87} potential.
The compact presentation, typically for a table, makes
it easy to grasp one important point: The more seriously and consistently
meson theory is pursued, the better the results.
This table and its trend towards the more comprehensive meson models
is the best proof for the validity of meson theory in the low-energy
nuclear regime.

While all models considered in table~1 describe the proton-proton ($pp$)
data well (with $\chi^2$/datum $\approx 2$), some models have a problem
with the neutron-proton ($np$) data (with $\chi^2$/datum $\approx 4-6$).
For the case of the Paris potential (and, in part, for the Nijmegen potential)
this is due to a bad
reproduction of the $np$ {\it total cross section} ($\sigma_{tot}$)
data.
When the latter data are ignored, the Paris potential fits $np$ as well as $pp$
(cf.\ table~1). The Nijmegen and the Paris potential predict the $np$
total cross sections too large because their $^3D_2$ phase shifts are
too large (cf.\ fig.~2b).

The $\chi^2$ given in table~1 are
calculated for the
range 10--300 MeV (in terms of the kinetic energy
of the incident nucleon in the laboratory system).
This energy intervall is very appropriate for reasons that we will explain
now.

For very low energies, the effective range expansion applies
and the quantitative nature of a potential can be tested by checking
how well it reproduces the empirical effective range
parameters.
We show this in table~2. It is seen that the Nijmegen and Paris
potential reproduce the four S-wave effective range parameters
very badly in terms of the $\chi^2$/datum, which is 95 and 169, respectively.
In the case of the Paris potential, this is entirely due to a wrong
prediction for $a^C_{pp}$, which is off by 26 standard deviations; the other
three parameters are described well.
For reasons of fairness, one should note that the
empirical value for $a^C_{pp}$ has a very small error
(namely, 0.0026 fm). This may blow up
small deviations by theory from experiment
in a misleading way. In absolute terms, the Paris $a^C_{pp}$ is wrong by
only 0.07 fm. This is negligible with regard to any
nuclear structure application.

The Nijmegen potential is off by about 10 standard deviations
for every parameter listed in table~2.
The large discrepancy between the triplet parameters
as predicted by the Nijmegen
potential and the empirical ones
is a reflection of the
fact that the
Nijmegen $^3S_1$  phase shifts are in general
 too low and too steep; e.~g., at 210
MeV, the Nijmegen potential predicts 14.91 degrees
for the $^3S_1$ phase shift
which is 12 standard deviations below the value from the
phase shift analysis
(which is $19.0\pm 0.33$ degrees~\cite{SAID}).

It is unreasonable to consider energies higher than 300 MeV
to test real NN potential models.
Around 285 MeV pion production starts. Thus, strictly speaking,
a real NN potential is inadequate above that energy.
It may be o.~k.\ to stretch this limit by a few MeV up to 300 MeV
(for practical reasons, since phase shift analyses often state their
results in steps of 100 MeV), but not beyond that. Also it is
quite consistent and
by no means accidental that
the $\chi^2$ for the fit of, particularly, the $np$ data becomes very
bad above 300 MeV for all potential models we are aware of.
For example, for the 567 neutron-proton data in the energy range
300--350 MeV the $\chi^2$/datum is 18.0, 7.2, and 6.4 for the
Nijmegen, Paris, and Bonn potential, respectively.
These $\chi^2$ are larger by about a factor of three as compared
to the ones for the data below 300 MeV (cf.\ table~1).

In the literatur, it is sometimes stressed that some real potentials are doing
particularly well for energies above 300 MeV. It is hard to understand
how it can be of any significance
when a real potential is doing well in an energy range where,
by principal, the real potential concept is inadequate.

In fig.~11, we give an overview of the fit of phase shifts by
some modern meson-theoretic potentials.
The solid line represents the prediction
by the Bonn full model~\cite{MHE87} while the dashed line is
the Paris~\cite{Lac80} prediction.
The Bonn full model is an energy-dependent potential. This energy-dependence
is inconvenient in nuclear structure applications. Therefore,
a representation of the model in terms of relativistic, energy-independent
Feynman amplitudes has been developed, using the relativistic,
three-dimensional Blankenbecler-Sugar method~\cite{BS66}.
This representation has become known as the
`Bonn B potential'~\cite{Mac89}.\footnote{See section 4 and appendix
A, table~A.1, of ref.~\cite{Mac89} for more details.}
The dotted line in fig.~11 shows the phase-shift predictions
by the Bonn B potential, which are
very similar to the ones by the Bonn full model.
For the Bonn~B potential, the $\chi^2$/datum is 2.1 for the 2158
$np$ data without $\sigma_{tot}$ and 2.3 for the 2322 $np$ data
including $\sigma_{tot}$ in the energy range 10--300 MeV (cf.\ table~1).
These numbers reflect well the fact that the Bonn~B potential describes the
the NN scattering data in close agreement with experiment
as well as with the predictions by the Bonn full model.
Moreover, since Bonn~B, like full Bonn, uses the full, relativistic
Feyman amplitudes, it is a non-local potential. Therefore,
it has a rather different off-shell behavior than local potentials
and, thus,
leads to different (but interesting)
results in nuclear structure (see section~4).

\section{Polemics}
Recently, there has been some debate about the `quality' of different
NN potentials. In particular, the $\chi^2$ of the fit of the experimental
NN data by a potential has sometimes become an issue.
Unfortunately, this debate has not always been conducted in a
strictly scientific manner. Therefore, we like to take
this opportunity for a few comments, in the hope that this may help
leading the discussion back to more scientific grounds.

{\it First: the $\chi^2$ is not a magic number.}

Its relevance with regard to the `quality' of a potential is
limited. Consider, for example,
a model based on little theory, but with many parameters;
this model will easily fit the data well and produce a very low
$\chi^2$ (e.~g., $\chi^2$/datum $\approx 1$).
But we will not learn much basic physics from this.
On the other hand, think of a model with
a solid theoretic basis and (therefore)
very few parameters (with each parameter having a physical meaning);
here, the comparison with the experimental data may teach us some
real physics. In such as case, a $\chi^2$/datum of 2 or 3 may be
excellent.

Thus, the $\chi^2$ represents only one aspect among several others
that need to be considered simultaneously
when judging the quality of a
NN potential. Other aspects of equal importance are the
theoretical basis of a potential model (see previous section)
and (closely related)
its off-shell behavior (that can, of course, not be tested by calculating the
$\chi^2$ with regard to the on-shell NN data).
This latter aspect is important, particularly, for the application
of a NN potential to nuclear structure.
In fact, in section~4 we will give an example, in which the
variation of the
$\chi^2$/datum between 1 and 6 affects nuclear struture result only in a
negligible way, while off-shell differences are of substantial
influence.

Notice also, that the $\chi^2$ sometimes blows up
small differences between theory and experiment
in a misleading way. This is so, in particular,
when the experimental error is very small (a good example
for this are the $pp$ data below 3 MeV, see discussion below).
In such cases, the $\chi^2$
is more a reflection of the experimental precision than of the
quality of the theory.

In summary, the $\chi^2$ must
be taken with a grain of salt;
overestimating the importance of the $\chi^2$ may
miss the physics.

{\it Second: if one considers the $\chi^2$,
it is insufficient to consider it for the $pp$ data only.}

In some recent work on $\chi^2$, persistently only $pp$ data are
considered~\cite{Swa89,Sto92}.
Since some potentials fit the $pp$ data much better than $np$
(cf.\ table~1), it may be tempting to do so (for reasons that, however,
have little to do with physics).
$pp$ states exist only for T=1 and, thus, a comparison with the $pp$
data tests only the T=1 potential.
Confrontation with the $np$ data tests both T=1 and T=0 and, thus,
is a much more comprehensive test of a potential.
In fact, the problems of some potentials
shown in
figs.~2 and 3 are due to T=0 states and, thus, are missed when only
$pp$ is considered.
Moreover, in nuclear structure calculations, typically both
the T=0 and the T=1 potential is needed.
In nuclear structure, the T=0 states are in general
as important as the T=1 states (in fact, one may well argue that the T=0
states are even more crucial since the very important
$^3S_1$ state is T=0).

{\it Third: if one calculates a $\chi^2$,
one should do it by all means properly.}

The most important rule here is:
{\it A $pp$ potential must only be confronted with
$pp$ data, while a $np$ potential must only be confronted with $np$
data.}
Though this rule is obvious, it has
sometimes been violated even by experts in the field.
Let us briefly explain why this rule is so important.
At low energies, NN scattering takes place mainly in $S$ wave.
There is well-known charge-dependence in the $^1S_0$ state and,
moreover,
the electromagnetic effects are very large
in low energy $pp$ scattering.
Thus, $np$ and $pp$ differ here substantially.
Moreover, there exist very accurate $pp$ cross section data
at low energies.
Consequently, if (improperly)
a $np$ potential is applied to $pp$ scattering,
a very large $\chi^2$ is obtained. However, this $\chi^2$
simply reflects the fact
that charge-dependence and Coulomb distorsion are important
and the $pp$ data carry a small error at low energy.

To give an example:
When  the $np$
versions of the Argonne~\cite{WSA84} and Bonn~\cite{MHE87} potentials
are (improperly)
confronted with the $pp$ data, a $\chi^2$/datum of 824 and 641,
respectively,
is obtained~\cite{Swa89}.
However, if (properly) the $pp$ version of the Bonn potential (see appendix)
is confronted with the $pp$ data, a $\chi^2$/datum of 1.9 is obtained
(cf.\ table~1).

Notice also that the change in the potential, that brings about
this enormous change in the $\chi^2$, is minimal.
The main effect comes from the $^1S_0$. A $np$ potential is fit
to the $np$ value for the singlet scattering length.
Now, if one wants to construct
a $pp$ potential from this, one has to do essentially only two things:
The Coulomb force has to be included and
the singlet scattering length
has to be readjusted to its $pp$ value. Since the scattering length
of an almost bound state is a super-sensitive quantity,
this is achieved by a very small
change of one of the fit parameters; for example, a change of the
$\sigma$ coupling constant by as little as 1\%.
This is all that needs to be done;
this
changes the
$\chi^2$/datum by several 100.

The difference in the $^1S_0$ phase shift
between $pp$ and $np$ is as small as the difference
between the solid and the dashed curve in the $^1S_0$ box of fig.~11
(which on the scale of the figure can hardly be seen).
When applied to nuclear matter, this difference
makes a change in the energy per nucleon of about 0.5 MeV
(of about 16 MeV total binding energy per nucleon).
Thus, we are dealing here with a very subtle difference, which by accident
makes a difference in the $\chi^2$/datum of 1.9 {\it versus} 600.
It shows in a clear way how misleading $\chi^2$ can be if
the reader is
not familiar with the field.

In any case, if one wants to test a $np$ potential, why not compare it
with the $np$ data? That's the most obvious thing to do; it's straightforward,
and nothing can be done wrong. Moreover,
this tests the T=0 and T=1 potential at once; thus, it is a comprehensive
and complete test of the quantitaive nature of a potential.
There is no need to bring by all means the $pp$ data into play
[except that for some potentials
the $pp$ $\chi^2$ is much better than the $np$ one (cf.\ table~1)].

Some potentials developed in the 1980's~\cite{WSA84,MHE87,Mac89}
are fit to the $np$ data, while older models, e.~g.\
refs.~\cite{Rei68,NRS78},
 were in general fit to $pp$.
The preference for fitting $np$ rather than $pp$ in recent years
has good reasons that we will explain now.
Nowadays,
phase shift analyses, see e.~g.\ ref.~\cite{Arn83,AHR87}, are conducted
in the following manner. First, the $pp$ data are analysed to determine
the $pp$ phase shifts. After this, the $np$ analysis is started.
However, this $np$ analysis is not performed independently, i.~e.\
it is not based upon the $np$ data only. The reason for this is that
there are twice as many partial waves for $np$ than for $pp$.
Therefore, a $np$ phase shift analysis
based upon the $np$ data only would have larger uncertainties.
Thus, for the T=1 $np$ phase shifts (except $^1S_0$),
the $pp$ analysis is used, after two corrections have been applied.
The Coulomb effect is removed and, in some cases, a simple assumption about
charge-dependence is applied (e.~g., the charge-dependence
of the one-pion-exchange due to pion-mass difference). Again,
this is done for $L>0$;
for $^1S_0$ an independent $np$ analysis is performed (confirming
the well-established charge-dependence in that state as seen
most clearly
in the well-known scattering length differences).
The result of the entire phase shift analysis is then stated in terms
of $np$ phase shifts~\cite{Arn83,AHR87}.

Now, from the way $pp$ and $np$ phase shift analyses
are conducted, it is clear that a potential that fits the
$np$ phase shifts well, will automatically fit the $pp$
phase shifts well, {\it after the small but important
corrections for Coulomb and charge-dependence have been applied}
(similarly to what is done in the phase shift analysis).
A good example for this
 is the Bonn potential that has originally been fit to $np$~\cite{MHE87}
and has a $\chi^2$/datum of 1.9 for the world $np$ data (cf.\ table~1).
Now, when the Coulomb force is included and
 the $^1S_0$ scattering length is adjusted to its $pp$ value,
then the world $pp$ data are reproduced with
a $\chi^2$/datum of also 1.9 (instead of 641~\cite{Swa89}, which is
a meaningless number).

In a more recent series of $\chi^2$ calculations~\cite{Sto92},
the Argonne and the Bonn $np$ potentials are again (improperly)
confronted with the $pp$ data. For the $pp$ data in the energy range
2--350 MeV a $\chi^2$/datum of 7.1 and 13, respectively, is obtained
(while for the range 0--350 MeV, the corresponding numbers are
824 and 641, as mentioned above).
However, to cut out the range 0--2 MeV is insufficient. Charge dependence
is important up to about 100 MeV. Thus, these new $\chi^2$
calculations~\cite{Sto92} are again meaningless.
Moreover, in ref.~\cite{Sto92} the ``Bonn 87'' model~\cite{MHE87}
is identified with a local $r$-space OBEP.
This is very incorrect and misleading.
In ref.~\cite{MHE87} the Bonn full model is presented with which,
therefore, the Bonn model of 1987 is to be identified.
Clearly, a Physics Reports article
published in 1987 does not serve the purpose to
present a simplistic model appropriate for the 1960's.
Since ref.~\cite{MHE87} is a review article, it contains
 also some discussion
of other (simpler) models.
In particular, comparison is made with the
local r-space concept (denoted by
OBEPR in ref.~\cite{MHE87})
to point out the
deficiencies of such simple models in fitting
certain phase shifts (similarly to what we explain in
section 2.1 of this contribution)
and to discuss to which extend such models
may still be usefull in some nuclear structure applications.
But clearly,
OBEPR is not `the Bonn model' and the $\chi^2$ of OBEPR is of no interest.

\section{NN potentials and nuclear structure}
One of the major topics of this workshop are nuclear structure calculations
based upon the bare NN interaction.
This program was started some 25 years ago by Kuo and Brown~\cite{KB66}.

For these miscroscopic nuclear structure calculations, a crucial question
is: what role do the differences between different NN potentials play?
We will try to answer this question in this section.
It is appropriate to subdivide the discussion into on-shell and
off-shell aspects.

\subsection{On-shell properties and nuclear structure}

By definition, we mean by ``on-shell properties of nuclear potentials''
their predictions for the deuteron and low-energy
two-nucleon scattering observables
($\lea 300$ MeV lab.\ energy).

To demonstrate the influence of on-shell differences on nuclear
structure results, one needs to find a case, where two potentials
are essentially identical off-shell (including the strength of
the tensor force, see discussion below), but differ on-shell.
It is not easy to isolate such a case, since, in general,
different potentials
show differences
on- and off-shell.

Recently, the Nijmegen group has constructed a revised
version~\cite{KSS92} of their original potential~\cite{NRS78}.
To improve the reproduction of the NN data, the new Nijmegen potential
fits the phase shifts of each NN partial wave separately
(allowing for a different set of fit parameters for each partial wave).
This is similar to the approach that Reid
took some 25 years ago~\cite{Rei68}
and, thus, this new potential is sometimes referred to
as the Nijmegen `Reid-like' potential~\cite{KSS92}.
In this approach, it is, of course, easy to obtain a perfect fit
of the NN scattering data.
Since both Nijmegen potentials are defined in terms of
the same mathematical expressions,
they have the same off-shell properties; in particular, they yield
the same D-state probability for the deuteron, implying equal
tensor force strength.
Thus, these two Nijmegen potentials represent an excellent example
for two potentials that are the same off-shell.
Now, on-shell these two potentials differ, since the more recent one
fits the NN data below 300 MeV much better than the older one.
This is seen most clearly in the $\chi^2$/datum for the fit of
the world $np$ data (including $\sigma_{tot}$)
which is improved from 6.5 for
the old Nijmegen potential to about 1.2 for the new potential.
Notice that this difference in the $\chi^2$
is substantial; probably the largest on-shell difference that exists
for any two realistic NN potentials.

{\it The question now is: how do these large on-shell differences
affect nuclear
structure results?}

To answer this question, it is best to consider a
nuclear structure quantity for
which an exact calculation can been performed.
We choose the binding energy of the triton.
Rigorous Faddeev calculations, which solve the three-body problem
exactly, are feasible and have actually been performed for the
two Nijmengen potentials under discussion. The results
are summarized in table~3:
We see that the old Nijmegen potential~\cite{NRS78}
with a $\chi^2$/datum of 6.5
predicts 7.63 MeV~\cite{FGP88} for the triton binding, while
the new potential with a $\chi^2$/datum of 1.2 yields
7.66 MeV~\cite{Fri91}.
Thus, in spite of the seemingly very large differences on-shell in terms
of the $\chi^2$, the difference in the nuclear structure quantity
under consideration is negligibly small.

On the other hand, consider two potentials that have an almost
identical $\chi^2$ for the world $np$ data (implying that they are
essentially identical on-shell). Accidentally, this is true for the
Paris~\cite{Lac80}
and the Bonn~B~\cite{Mac89} potential, which both have a $\chi^2$/datum
of about 2 (cf.\ table~3).
The triton binding energy predictions derived from these two potentials
are 7.46 MeV for the Paris potential and 8.13 MeV for Bonn~B.
This appears like a contradiction to the previous  results.
With the $\chi^2$ so close, naively,
one would have expected identical triton binding energy predictions.
Obviously there is another factor, even more important
than the $\chi^2$: this factor is the off-shell behavior of a potential
(particularly, the off-shell tensor force strength, that can vary
substantially for different realistic potentials). A simple
measure for this strength is the D-state probabilty
of the deuteron, $P_D$, with a smaller $P_D$ implying a weaker
(off-shell) tensor force. For this reason, we are also giving in table~3
the $P_D$ for each of the potentials under consideration.
The real reason for the differences in the predictions
by Paris and Bonn~B is the difference in the $P_D$ with Paris predicting
5.8\% and Bonn~B 5.0\% (cf.\ table~3).

Summarizing:
on-shell differences between potentials are seen in differences
in the $\chi^2$ for the fit of the NN data; off-shell differences are
seen in differences in $P_D$. As table~3 reveals,
off-shell differences are much more important
than on-shell differences for the triton binding energy.\footnote{Here
and in the following, it is understood that we consider only realistic
potentials that yield a reasonable description of the empirical
NN data. Arbitrary on-shell variations may, of course, have large
effects on nuclear structure predictions.}

A similar consideration can be done for nuclear matter.
This is summarized
in table~4.
In this example, we are more specific as far as the $\chi^2$ is concerned.
We have choosen a particular set of NN data, namely the
$np$ analysing power ($A_y$)
data at 25 MeV, which have been measured with great accuracy~\cite{Sro86}.
This experimental data set tests, in particular, the triplet P-waves.
Note, that differences between modern potentials occur mainly in these
P-waves (besides the ones in D-waves shown in fig.~2).
We consider now
two potentials which fit these data perfectly, namely
Paris and Bonn~B ($\chi^2$/datum around 1 in both cases) and another
potential with the rather poor $\chi^2$/datum of 3.3
(denoted by `Potential $B'$').
These three potentials are applied to nuclear matter in a standard
Brueckner calculation. It turns out (see table~4) that
the large $\chi^2$ difference between Bonn~B and Potential $B'$
makes only a difference of 0.3 MeV in the
nuclear matter energy.
On the other hand, the nuclear matter predictions by the other
two potentials (with
identical and perfect fit)
differ by as much as 1.4 MeV.
As in the case of the triton,
the difference in $P_D$ is the reason.
[Notice that this difference shows up in the $^3S_1$ contribution
(cf.\ table~4)].

Again, the off-shell behavior turns out to be of great relevance.
We will now discuss these off-shell aspects in more detail.

\subsection{Off-shell effects}
For a given NN potential $V$, the $T$-matrix for free-space
two-nucleon scattering is obtained from the Lippmann-Schwinger
equation, which reads in the center-of-mass system
\begin{equation}
T({\bf q'}, {\bf q}) = V({\bf q'}, {\bf q}) -
\int d^3k V({\bf q'}, {\bf k}) \frac{M}{k^2-q^2-i\epsilon}
T({\bf k}, {\bf q})
\end{equation}
and in partial-wave decomposition
\begin{equation}
T^{JST}_{L'L}({ q'}, { q}) = V^{JST}_{L'L}({ q'}, { q}) -
\sum_{L''} \int_0^\infty k^2 dk V^{JST}_{L'L''}({ q'}, { k})
\frac{M}{k^2-q^2-i\epsilon}
T^{JST}_{L''L}({ k}, { q})
\end{equation}
where $J,S,T$, and $L$ denote the total angular momentum, spin, isospin,
and orbital angular momentum, respectively, of the two nucleons; and
$M$ is the mass of the free nucleon.
{\bf q}, {\bf k}, and ${\bf q'}$ are the initial, intermediate,
 and final relative three-momenta, respectively
($q'\equiv |{\bf q'}|, k\equiv |{\bf k}|, q\equiv |{\bf q}|$).

Notice that the integration over the intermediate
momenta $k$ in eqs.~(1) and (2) extends from
zero to infinity. For intermediate states with $k\neq q$, energy
is not conserved and the nucleons are off their energy shell (`off-shell').
The off-shell part of the potential (and the $T$-matrix) is involved.
Thus, in the integral term in eqs.~(1) and (2),
the potential (and the $T$-matrix) contributes essentially off-shell.

However, the off-shell potential does not really play any role
in free-space NN scattering.
The reason for this is simply the procedure by which NN potentials
are constructed. The parameters of NN potentials are adjusted such
that the resulting on-shell $T$-matrix fits the empirical
NN data.
For our later discussion, it is important to understand this point.
Let us consider a case in which the off-shell contributions
are particularly large, namely the on-shell $T$-matrix
in the $^3S_1$ state:
\begin{eqnarray}
T^{110}_{00}({ q}, { q}) & = & V^{110}_{00}({ q}, { q}) -
 \int_0^\infty k^2 dk V^{110}_{00}({ q}, { k})
\frac{M}{k^2-q^2-i\epsilon}
T^{110}_{00}({ k}, { q}) \nonumber \\
     &    & \mbox{}
- \int_0^\infty k^2 dk V^{110}_{02}({ q}, { k})
\frac{M}{k^2-q^2-i\epsilon}
T^{110}_{20}({ k}, { q})
\end{eqnarray}
Up to second order in $V$, this is
\begin{eqnarray}
T^{110}_{00}({ q}, { q}) & \approx &
V^{110}_{00}({ q}, { q}) -
 \int_0^\infty k^2 dk V^{110}_{00}({ q}, { k})
\frac{M}{k^2-q^2-i\epsilon}
V^{110}_{00}({ k}, { q}) \nonumber \\
     &    & \mbox{}
- \int_0^\infty k^2 dk V^{110}_{02}({ q}, { k})
\frac{M}{k^2-q^2-i\epsilon}
V^{110}_{20}({ k}, { q}) \\
    & \approx &
V^{110}_{00}({ q}, { q})
- \int_0^\infty k^2 dk V^{110}_{02}({ q}, { k})
\frac{M}{k^2-q^2-i\epsilon}
V^{110}_{20}({ k}, { q})
\end{eqnarray}
where in the last equation, we also neglected the second order in
$V^{110}_{00}$ which is, in general, much smaller than the second order
in $V^{110}_{02}$.
Or, without partial-wave decomposition
\begin{equation}
T({\bf q}, {\bf q}) \approx V_C({\bf q}, {\bf q}) -
\int d^3k V_T({\bf q}, {\bf k}) \frac{M}{k^2-q^2-i\epsilon}
V_T({\bf k}, {\bf q})
\end{equation}
where $V_C$ denotes the central force and $V_T$ the tensor force.
The tensor force makes possible transitions between states
that are not diagonal in $L$.

The on-shell $T$-matrix is related to the observables that are measured
in experiment.
Thus, potentials which fit the same NN scattering data
produce the same on-shell $T$-matrices.
However, this does not imply that the potentials are the same.
As seen in eqs.~(1) and (2), the $T$-matrix is the sum of two terms, the Born
term and an integral term. When this sum is the same, the individual terms
may still be quite different.

Let's consider an example. The $T$-matrix in the $^3S_1$ state is attractive
below 300 MeV lab.\ energy.
If a potential has a strong (weak) tensor force, then the integral term
in eq.~(5) is large (small), and the negative Born term will be small (large)
to yield the correct on-shell $T$-matrix element.

In fig.~12 we show the $^3S_1-^3D_1$ potential matrix element,
$V^{110}_{02}(q,k)$, for the Paris and the Bonn~B potential.
The momentum $q$ is held fixed at 153 MeV which is equivalent
to a lab.\ energy of 50 MeV ($E_{lab}=2q^2/M$).
The abscissa, $k$, is the variable over which the integration
in eq.~(5) is performed.
It is seen that, particularly for large off-shell momenta,
the Bonn~B potential
is smaller than the Paris potential.
However, notice also that at the on-shell point ($q=k$, solid dot in fig.~12)
both potentials are identical (both potentials predict the
same $\epsilon_1$ parameter).
Thus, the Bonn~B potential has a weaker {\it off-shell}
tensor force than the Paris
potential. Since the Bonn~B and the Paris potential predict almost
identical phase shifts~\footnote{For
the $^3S_1$ phase shift at 50 MeV,
Paris predicts 62.3$^0$ and Bonn-B 62.2$^0$},
the Born term (central force) in the $^3S_1$
state will be more attractive for the Bonn~B potential than for the
Paris potential.

In summary, fig.~12 demonstrates in a clear way
how large off-shell differences can be between
two realistic potentials that produce identical phase parameters
(identical on-shell $T$-matrices).

A neat measure  of the strength
of the nuclear tensor force is the $D$-state probability
of the deuteron, $P_D$. This quantity is defined as
\begin{equation}
P_D=\int_0^\infty q^2 dq \psi^2_2 (q)
\end{equation}
where $\psi_2$ denotes the deuteron $D$-wave in momentum space.
The deuteron waves are obtained from a system of coupled
integral equations
\begin{equation}
\psi_L (q) = -\frac{1}{B_d + q^2/M}
\sum_{L'} \int_0^\infty k^2 dk V^{110}_{LL'}(q,k) \psi_{L'} (k)
\end{equation}
where $B_d$ denotes the binding energy of the deuteron.
Approximately, we can write for the $D$-wave
\begin{equation}
\psi_2 (q) \approx -\frac{1}{B_d + q^2/M}
\int_0^\infty k^2 dk V^{110}_{20} (q,k) \psi_0 (k)
\end{equation}
where $\psi_0$ denotes the deuteron $S$-wave.
Notice that the off-shell potential
shown in fig.~12 is integrated over here.
This explains why the D-state probability for Bonn B is smaller than for Paris,
in spite of both potentials having the same on-shell properties.

In the Brueckner approach to the nuclear many-body problem,
a $G$-matrix is calculated which is the solution of the
equation
\begin{equation}
G({\bf q'}, {\bf q}) = V({\bf q'}, {\bf q}) -
\int d^3k V({\bf q'}, {\bf k}) \frac{M^\star Q}{k^2-q^2}
G({\bf k}, {\bf q})
\end{equation}
The similarity to the Lippmann-Schwinger equation, eq.~(1), is obvious.
There are differences in two points:
The Pauli projector $Q$ and the energy denominator.
(For simplicity, we have assumed the so-called continuous choice
for the single particle energies in nuclear matter, i.~e., the energy
of a nucleon is represented by $\epsilon(p) = p^2/(2M^\star) - U_0$
for $p\leq k_F$ as well as $p> k_F$, where $M^\star$ is the effective
mass and $U_0$ a constant.)
The Pauli projector prevents scattering into occupied states and, thus,
cuts out the low momenta in the $k$ integration.
The difference introduced by the Pauli projector is known as the Pauli effect,
the energy denominator gives rise to the so-called dispersive effect.
When using the continuous choice, the dispersive effect is given simply
by the replacement of $M$ by $M^\star$ ($\approx 2/3 M$ at nuclear
matter density) in the numerator of the integral
term, which simply leads to a reduction of this attractive term
by a factor $M^\star/M$.
Both effects go into the same direction, namely they quenche the
integral term. Since the integral term is negative, these effects
are repulsive.

 From our previous discussion, we know already that a potential with a weak
tensor force (small $P_D$) produces a smaller integral term than
a strong tensor-force potential. Thus, the $G$-matrix resulting
from a strong-tensor force
potential will be subject to a larger quenching;
thus, the $G$-matrix will be less attractive than the one
produced by a weak-tensor force potential.
This explains why NN interactions with a weaker tensor force
yield more attractive results when applied to nuclear few- and
many-body systems.

In the case of the three-nucleon system, the formalism is different,
but the mechnisms at work are very similar. A two-nucleon $T$-matrix is used
as input for the Faddeev equations. The energy parameter of this $T$-matrix
runs from --8.5 MeV to $-\infty$, causing a large dispersive effect
(there is no Pauli effect).

We show now three examples
which demonstrate clearly these differences in predictions
for nuclear energies by weak tensor-force potentials
{\it versus} strong tensor-force potentials.
As examples we choose
 the binding energy of the triton (fig.~13),
the spectrum of a s-d shell nucleus ($^{21}$Ne, fig.~14), and
nuclear matter (fig.~15).
To obtain an idea of the strength of the tensor-force component
of the various potentials applied, we list
$P_D$ in table~5.\footnote{The Bonn~A potential that occurs in
some of the figures is a variation of Bonn~B with an even weaker
tensor force, but otherwise very similar to Bonn~B. Predictions by
Bonn~A are slightly more attractive as compared to Bonn~B, but the difference
is not substantial; e.~g., for the triton binding energy Bonn~B
predicts 8.1 MeV while Bonn~A yields 8.3 MeV.}

The three examples suggest that potentials with weak tensor force
(as implied by relativistic meson theory)
may be superior in explaining nuclear structure phenomena.

\section{Summary and conclusions}
We have described the developments in the field of realistic
NN interactions since the event of the Kuo-Brown matrix elements
in 1966. It turns out that each of three models for the NN interaction
currently in use, represents one decade of the past 30
years.
The Nijmegen potential~\cite{NRS78} is an excellent example for
approaches typical for the 1960's, the Paris potential
a representative of the 1970's, and the Bonn full model
for the 1980's.
Moreover, we could clearly reveal that with
the development of more consistent and comprehensive
meson-models over that period of time
the quantitative explanation of the NN data has continuously improved.
This fact is one of the simplest and best arguments for the
appropriateness of
meson models in low energy nuclear physics.

We have also taken this opportunity to correct some misconceptions
concerning the question of how to test the quantitative nature of NN
potentials. Of particular importance is: {\it predictions by
pp potentials must only be compared with pp data, while predictions
by np potentials must only be compared with np data.}
Following this rule, the $\chi^2$/datum for the fit of the relevant
world NN data (as obtained by independent researchers in the
field~\cite{SAID}) comes out to be 5.1, 3.7, and 1.9 for the
Nijmegen~\cite{NRS78}, Paris~\cite{Lac80}, and Bonn~\cite{MHE87}
 potential, respectively (cf.\ table~1).

Furthermore, we have investigated the influence that differences between
different NN potentials have on nuclear structure predictions.
It turns out that for potentials that fit the NN data reasonably well,
on-shell differences have only a negligible  effect.
However, potentials that are essentially identical on-shell,
may differ substantially off-shell.
Such off-shell differences may lead to large differences
in nuclear structure predictions.
Relativistic, meson-theory based
potentials (which are non-local)
are in general weaker off-shell than their
local counterparts.
In particular, the weaker (off-shell) tensor force component
(as quantified by a small deuteron D-state probability, $P_D$)
leads to more binding in finite nuclei. For several examples
shown, these predictions compare favourably with experiment.

Happy Birthday to Tom Kuo! Our present for Tom is fig.~14 which
shows that the predictive power of the Kuo-Brown approach is even better
than anticipated some 25 years ago; last not least, this is due to
progress in our
understanding of the nuclear force.
\\
\\
\\
{\bf Acknowledgement.}
It is a pleasure to thank F. Sammarruca for suggestions on the
manuscript.
This work was supported in part by the U.S. National Science Foundation
under Grant Nos.~PHY-8911040 and PHY-9211607,
and by the Idaho State Board of Education.

\newpage

\appendix
\section{\hspace*{-.7cm}ppendix:
The proton-proton phase-shifts of the Bonn full model}
In the basic paper about the Bonn full model~\cite{MHE87},
only neutron-proton ($np$) phase shifts are listed (table~2
of ref.~\cite{MHE87}).
The reason for this is the fact that modern
phase-shift analyses~\cite{Arn83,AHR87} state their results
in terms of $np$ phase shifts---for practical reasons:
the $np$ system exists for all (T=0 and T=1)
partial waves and, thus, with $np$ one can cover all cases.
In contrast, $pp$
is restricted to the T=1 two-nucleon states.
As explained in detail in section~3, from the way phase-shift analyses are
conducted it is clear that a potential that fits the $np$ phase shifts
well, will
automatically also fit the $pp$ phase shifts well, {\it
after two minor (but obvious) adjustments
are done for the $pp$ case:}
inclusion of the Coulomb force and
fit of the $pp$ singlet scattering length (instead of its $np$ value).
Therefore, a potential which has been tested successfully in $np$ does
not necessarily have to be tested also in $pp$.
The successfull description of also the $pp$ data can be anticipated.

To prove this point, one has to calculate
the  $pp$ phase shifts correctly.
In this appendix, we will do this for the Bonn full
model~\cite{MHE87}. Since a first calculation~\cite{HH89} of the Bonn $pp$
phase shifts contains some errors,
let us stress the
important points of the $pp$ phase shift calculation:
\begin{enumerate}
\item
In the scattering equation, the correct proton mass of 938.272 MeV
is used
(and not the average nucleon mass of 938.926 MeV appropriate for $np$).
\item
The improved, relativistic Coulomb potential~\cite{Bre55,AS83} is
applied~\footnote{See eqs.~(25), (26), and (10) of
ref.~\cite{Ber88}; in the notation of ref.~\cite{Ber88}, we
use $V_{C1}$ for the Coulomb potential.}
 (and not the static
Coulomb potential).~\footnote{The
authors like to thank
Dr.\ Vincent Stoks for drawing their attention to the improved
Coulomb potential of ref.~\cite{AS83}.}
\item
The $^1S_0$ scattering length is fit to its experimental
$pp$ value of $a^C_{pp}=-7.8197$ fm (cf.\ table~2) by changing the coupling
constant of the $\sigma'$ contribution to 5.6037 (for $np$
scattering, 5.6893 is used~\cite{MHE87}).
The effective range parameter comes out to be $r^C_{pp}=2.7854$ fm,
in good agreement with the empirical value (cf.\ table~2).
This refit is used only for $^1S_0$; the other $pp$ partial waves are kept
unchanged as compared to the $np$ case, except for the Coulomb effect.
This is the most reasonable assumption, unless one derives
all charge-dependent effects accurately from the underlying meson model,
which is not our intention here (it is an interesting
and comprehensive project for the future).
\end{enumerate}

The $pp$ phase shifts obtained from this calculation
(in which the momentum-space method of
ref.~\cite{VP74} is applied\footnote{The authors
are indebted to Dr.\ J. Haidenbauer
for some codes involved in the momentum-space Coulomb problem.})
are listed in table~6.
Calculating the $pp$ observables from these phase shifts and comparing
with the world $pp$ data, yields a $\chi^2$/datum of 1.94 (cf.\
table~1). The $\chi^2$/datum for the corresponding world $np$ data
is 1.87 (using the $np$ version of the Bonn full model).
This confirms our statement that a potential that fits $np$
well will
automatically also fit $pp$ well {\it (if the proper $pp$ phase
shifts are used)}.

\pagebreak
{\bf Table~1.} Comparison of some meson-theoretic nucleon-nucleon
potentials.
\begin{center}
\begin{tabular}{lccc}
\hline\hline
              & Nijmegen~\cite{NRS78}
& Paris~\cite{Lac80} & Bonn full model~\cite{MHE87} \\
\hline
{\bf \# of free parameters}
                     &    15 & $\approx 60$   & 12 \\
{\bf Theory includes:}\\
OBE terms     &     Yes        &  Yes  &  Yes     \\
$2\pi$ exchange&    No         &  Yes  &  Yes      \\
$\pi\rho$ diagrams& No         &  No   &  Yes     \\
Relativity   &      No         &  No   &  Yes      \\
\multicolumn{4}{l}{\bf\boldmath $\chi^2$/datum for fit of world NN data$^a$:}\\
all $pp$ data    & 2.06             &  2.31  &  1.94   \\
all $np$ data    &6.53              &  4.35  &  1.88   \\
($np$ without $\sigma_{tot}$)&(3.83)& (1.98) & (1.89)  \\
all $pp$ and $np$&5.12              &  3.71  &  1.90   \\
\hline\hline
\end{tabular}
\end{center}
$^a$ The $\chi^2$/datum  are obtained
from the computer software SAID of R. A. Arndt and L. D. Roper
(VPI\&SU)~\cite{SAID}.
The world NN data set in the range 10--300 MeV as of September 1992
is used; it includes 1070 data for $pp$, 2158 data
for $np$ without total cross sections
($\sigma_{tot}$), and 2322 data for $np$ with $\sigma_{tot}$.
\normalsize

\newpage

{\bf Table~2.} Low-energy $S$-wave scattering parameters
as predicted by some
meson-theoretic NN potentials. In the last row, the $\chi^2$/datum
is given for the fit of
the 4 empirical data given in the
last column.
\begin{center}
\begin{tabular}{ccccc}
\hline\hline
               & Nijmegen~\cite{NRS78} & Paris~\cite{Lac80}
& Bonn full model~\cite{MHE87} & Empirical\\
\hline
{\bf Singlet:}\\
$a^C_{pp}$ (fm) & --7.797 & --7.887$^a$ & --7.8197$^b$ & --7.8196(26)$^c$\\
$r^C_{pp}$ (fm) & 2.697   & 2.805$^a$   & 2.7854$^b$   & 2.790(14)$^c$\\
{\bf Triplet:}\\
$a$ (fm)         & 5.468 & 5.427 & 5.427 & 5.424(4)$^d$\\
$r$ (fm)         & 1.818 & 1.766 & 1.755 & 1.759(5)$^d$\\
{\bf {\boldmath $\chi^2$}/datum}& 95.0 & 168.9 & 0.33 & --- \\
\hline\hline
\end{tabular}
\end{center}
$^a$ The predictions by the Paris potential for $a^C_{pp}$ and $r^C_{pp}$
given in this table were obtained in independent calculations
 by Piepke~\cite{Pie85} and
the Nijmegen group~\cite{Ber88}; they differ from ref.~\cite{Lac80}.\\
$^b$ See appendix.\\
$^c$ From ref.~\cite{Ber88}.\\
$^d$ From ref.~\cite{Dum83}.
\normalsize

\pagebreak

{\bf Table~3.} Correlations between two-nucleon and
three-nucleon properties as predicted by
different NN potentials.
The $\chi^2$/datum is related to the on-shell properties of a potential,
while $P_D$ depends essentially on the off-shell behavior.
\begin{center}
\small
\begin{tabular}{lcccc}
\hline\hline
             & Nijmegen~\cite{NRS78}& Nijm.-Reid~\cite{KSS92}
 & Bonn~B~\cite{Mac89} &    Paris~\cite{Lac80}\\
\hline
{\bf Two-nucleon data:}\\
$\chi^2$/datum$^a$ & 3.8    & 1.2    &  2.1   &   2.0\\
$P_D$ (\%) $^b$&     5.4    & 5.4    &  5.0   &   5.8\\
{\bf Triton binding (MeV):}& 7.63 & 7.66 & 8.13 & 7.46\\
\hline\hline
\end{tabular}
\end{center}
$^a$ for the fit of the world $np$ data (without $\sigma_{tot}$)
in the range 10--300 MeV (cf.~table~1).\\
$^b$ $D$-state probability of the deuteron.\\

\pagebreak
{\bf Table~4.} Correlations between two-nucleon and nuclear matter properties
for different potentials. Nuclear matter
with a Fermi momentum of $k_F=1.35$ fm$^{-1}$ is considered.
\begin{center}
\begin{tabular}{lccc}
\hline\hline
             & Bonn~B~\cite{Mac89}& Potential $B'$    &    Paris~\cite{Lac80}\\
\hline
\multicolumn{4}{l}{\bf Two-nucleon data:}\\
$\chi^2$/datum$^a$ & 1.0          &  3.3            &    1.1\\
$P_D$ (\%) $^b$&     5.0          &    5.0          &     5.8\\
\multicolumn{4}{l}{\bf Nuclear matter energy (MeV):}\\
$^3S_1$ contribution &   --18.8         &  --18.8         &    --17.1\\
Total         &    --12.1         &  --11.8         &    --10.7\\
Difference to Bonn~B&  ---        &    0.3          &   1.4\\
\hline\hline
\end{tabular}
\end{center}
$^a$ for the fit of the $np$ analysing power data at 25 MeV
(16 data points)~\cite{Sro86}.\\
$^b$ $D$-state probability of the deuteron.\\

\pagebreak

{\bf Table~5.} $D$-state probaility of the deuteron, $P_D$,
 as predicted by various
potential applied in figs.~13--15.
\begin{center}
\begin{tabular}{lc}
\hline\hline
Potential         &     $P_D$ (\%) \\
\hline
Bonn~A~\cite{Mac89} & 4.4\\
Bonn~B~\cite{Mac89} & 5.0\\
Bonn~C~\cite{Mac89} & 5.6\\
Paris~\cite{Lac80}  & 5.8\\
TRS~\cite{TRS75}    & 5.9\\
Argonne $V_{14}$~\cite{WSA84} & 6.1\\
Reid (RSC)~\cite{Rei68} & 6.5\\
\hline\hline
\end{tabular}
\end{center}

\pagebreak

{\bf Table~6.} Proton-proton phase shifts (in degrees) as predited by the
Bonn full model~\cite{MHE87}.
\begin{center}
\begin{tabular}{ccccccc}
\hline\hline
$E_{lab}$ (MeV) &  25 &   50   &  100   &  150   &  200  & 300   \\
\hline
$^1S_0$     &  48.309 & 38.315 & 23.860 & 13.298 & 4.823 & --8.288\\
$^3P_0$     &   9.209 & 12.633 & 10.978 &  6.223 & 0.923 & --9.349\\
$^3P_1$     &  --4.954 & --8.295 & --13.120&--17.327 &--21.406&--29.506\\
$^1D_2$     &   0.692 &  1.677 &   3.700 & 5.555 &  6.998&  8.334\\
$^3P_2$     &   2.371 &  5.662 &  10.945 & 14.117 & 15.866& 17.238\\
$^3F_2$     &   0.104 & 0.339  & 0.804   &  1.138 & 1.282 & 0.900\\
$\epsilon_2$ & --0.822 & --1.751 & --2.743 & --2.987 & --2.875 & --2.276\\
$^3F_3$     &  --0.230 & --0.700 & --1.564 & --2.199 & --2.662 & --3.318\\
$^1G_4$     &  0.039  &  0.153 &  0.421 &  0.690 &  0.967 & 1.561\\
$^3F_4$     &  0.019  &  0.100 &  0.415 & 0.860 &  1.366 & 2.393\\
$^3H_4$     &    0.004 & 0.025 & 0.108 &   0.214 & 0.327 & 0.539\\
$\epsilon_4$ & --0.048 & --0.195 & --0.548 & --0.863 & --1.122 & --1.482
\\ \hline\hline
\end{tabular}
\end{center}
This table lists only a small selection of partial-waves and energies.
The $pp$ phase-shifts of the Bonn full model for any energy in the range
10--325 MeV and any partial wave with $J\leq 7$ can be obtained from
SAID~\cite{SAID}.

\pagebreak

\begin{center}
\large Figure Captions
\end{center}

\hspace*{-.6cm}{\bf Figure 1.}
One-boson-exchange (OBE) model for the NN interaction.
\\
{\bf Figure 2.} Phase-shifts of NN scattering for the {\bf (a)} $^1P_1$,
{\bf (b)} $^3D_2$, and {\bf (c)} $^3D_3$ partial wave. Predictions are shown
for
the Bryan-Scott
(B-S) potential of 1969~\cite{BS69} (long dashes),
Nijmegen potential~\cite{NRS78} (short dashes),
Paris potential~\cite{Lac80} (dotted),
and the Bonn full model~\cite{MHE87} (solid line).
The solid squares represent the energy-independent phase shift analysis by
Arndt {\it et al.}~\cite{AHR87}.
\\
{\bf Figure 3.} {\bf (a)} Neutron-proton spin correlation parameter $C_{NN}$
at 181 MeV.
 Predictions by the Nijmegen potential~\cite{NRS78} (long dashes),
the Paris potential~\cite{Lac80} (dotted), and
the Bonn full model~\cite{MHE87} (solid line)
are compared with the data (solid squares)
from Indiana~\cite{Indiana}.
 The $\chi^2/$datum for the fit of these data is
54.4 for Nijmegen, 3.22 for Paris, and 1.78 for Bonn.
The experimental error bars include only systematics and statistics;
there is also a scale error of $\pm 8$\%.
In the calculations of the $\chi^2$, all three error have been
taken into account~\cite{SAID}.\\
{\bf (b)} Same as (a), but at 220 MeV with the data from TRIUMF~\cite{TRIUMF}.
 The $\chi^2/$datum for the fit of these data is
121.0 for the Nijmegen, 16.1 for the Paris, and 0.49 for
the Bonn B potential~\cite{Mac89}.
In addition to the experimental error shown, there is a scale uncertainty
of $\pm 5.5$\%. In the calculation of the $\chi^2$,
all errors were taken into account~\cite{SAID}.
\\
{\bf Figure 4.} The $2 \pi$-exchange contribution to the NN interaction
as viewed by dispersion theory. Solid lines represent nucleons, dashed lines
pions.
\\
{\bf Figure 5.} Two components of the Paris potential:
{\bf (a)} the singlet central potential and
{\bf (b)} the (T=1, S=1) tensor potential.
The dotted line represents the result as derived from the
underlying theory. The solid line is the parametrized Paris
best-fit potential~\cite{Lac80}.
\\
{\bf Figure 6.} Field-theoretic model for the $2 \pi$-exchange.
Solid lines represent nucleons, double lines isobars, and dashed lines pions.
The hatched circles are $\pi \pi$ correlations.
\\
{\bf  Figure 7.} Phase shifts of some peripheral partial waves
as predicted by a field-theoretic model for the $2\pi$
exchange (solid line, ``BONN''~\cite{MHE87}) and
by dispersion theory (dotted line labeled ``P'73''~\cite{Vin73}).
Both calculations also include OPE and one-$\omega$ exchange.
The dotted line labeled ``P'80'' is the fit by the parametrized Paris
potential~\cite{Lac80}. Octagons represent the phase shift analysis
by Arndt {\it et al.}~\cite{Arn83} and triangles the one by Bugg and
coworkers~\cite{Dub82}.
\\
{\bf Figure 8.} $\pi \rho$ contributions {\it versus} phenomenology
in {\bf (a)} $^1D_2$,
{\bf (b)} $^3D_2$, and {\bf (c)} $^3D_3$.
The curves labeled `$2 \pi$ Paris' and `$2 \pi$ Bonn'
represent the predictions by the Paris and Bonn model, respectively,
when only the contributions from $\pi$, $2 \pi$, and $\omega$ are
taken into account.
Adding the phenomenological short-range potential
yields the dotted `Paris' curve (parametrized Paris
potential~\cite{Lac80}). Adding the $\pi \rho$ contributions
(Fig.~9) yields the solid `$2\pi+\pi\rho$ Bonn' curve
(Bonn full model~\cite{MHE87}).
\\
{\bf Figure 9.} $\pi \rho$ contributions to the NN interaction.
\\
{\bf Figure 10.} Effect of the $\pi\rho$ diagrams in some S- and P-waves.
The dashed curve is the prediction by the $\pi+2\pi+\omega$
model. The solid line is obtained when the $\pi\rho$ contributions
are added.
\\
{\bf Figure 11.} Overview of the fit of phase shifts by some current models
for the NN interaction.
Predictions are shown for the Bonn full model~\cite{MHE87} (solid line),
the Paris
potential~\cite{Lac80} (dashed),
 and the Bonn B potential~\cite{Mac89} (dotted).
Octagons represent the phase shift analysis by
Arndt {\it et al.}~\cite{Arn83},
triangles the one by the Dubois {\it et al.}~\cite{Dub82},
and the solid dot is from the recent analysis by Bugg~\cite{Bug90}.
\\
{\bf Figure 12.} Half off-shell potential $<^3\!S_1|V(q,k)|^3D_1>$.
$q$ is held fixed at 153 MeV. The solid curve is the Bonn~B
potential~\cite{Mac89} and the dashed curve the Paris
 potential~\cite{Lac80}.
The solid dot denotes the on-shell point ($k=q$).
\\
{\bf Figure 13.} Triton energy, ${\cal E}_t$,
{\it versus} the invers charge radius
of $^3$He, $1/r_c$, as predicted by various NN potentials.
The experimental value is given by the horizontal error bar.
\\
{\bf Figure 14.} The spectrum of $^{21}$Ne. Predictions by
NN potentials are compared with experiment. (From ref.~\cite{Jia92}.)
\\
{\bf Figure 15.} Energy per nucleon in nuclear matter, ${\cal E}/A$,
{\it versus} density expressed in terms of the Fermim momentum
$k_F$. Dashed lines represent results from non-relativistic Brueckner
calculations, while solid lines are Dirac-Brueckner results.
The letters A, B, and C refer to the Bonn A, B, and C potential,
respectively. The shaded square covers empirical information
on nuclear saturation. Symbols in the background denote saturation
points obtained for a variety of NN potentials applied in conventional
many-body theory. (From ref.~\cite{BM90}.)

\end{document}